\documentclass[preprint,showpacs,preprintnumbers,amsmath,amssymb]{revtex4}

\usepackage{graphicx}
\usepackage{dcolumn}
\usepackage{bm}

\begin{document}


\title{Vector order parameter in general relativity.\\ Covariant equations}

\author{Boris E. Meierovich}

\affiliation{P.L.Kapitza Institute for Physical Problems.  \\ 2 Kosygina street, Moscow 119334, Russia}
\homepage{http://www.kapitza.ras.ru/people/meierovich/Welcome.html}

\date{\today}

\begin{abstract}
Phase transitions with spontaneous symmetry breaking and vector order parameter are considered in multidimensional theory of general relativity. Covariant equations, describing the gravitational properties of topological defects, are derived. The topological defects are classified in accordance with the symmetry of the covariant derivative of the vector order parameter. The abilities of the derived equations are demonstrated in application to the brane world concept. New solutions of the Einstein equations with a transverse vector order parameter are presented. In the vicinity of phase transition the solutions are found analytically.
\end{abstract}

\pacs{04.50.-h, 98.80.Cq}
\maketitle

\section{\label{sec1}Introduction}
	
Macroscopic Landau theory of phase transitions with changes of symmetry was initially developed for crystals \cite{Landau}. Cosmological implication of phase transitions was initiated by Kirzhnits \cite{Kirzhnits}, and modern standard cosmology is actually a sequence of phase transitions with spontaneous symmetry breaking (grand unification, electro-weak, quark-hadron, ...). The macroscopic theory of phase transitions allows to consider astrophysical phenomena self-consistently, even without the knowledge of the nature of physical vacuum.
	
The theories of brane world and multidimensional gravity continue numerous attempts to apply the spontaneous symmetry breaking for clarifying the origin of long-standing problems in physics, such as enormous hierarchy of energy and mass scales, dark matter and dark energy effects, and others. In particular, the brane world is considered as a topological defect, inevitably accompanying the phase transitions with spontaneous symmetry breaking. Topological defects (domain walls, strings, monopoles, textures,...) are described mathematically by the order parameter, appearing in the state of lower symmetry.
	
Following the pioneer works of Polyakov \cite{Polyakov} and 't Hooft \cite{Hooft}, the topological defects were generally described by single scalar fields, or by the multiple sets of scalar fields. In brane world theories the hedgehog-type configurations of scalar multiplets in extra dimensions played the role of the order parameters (see \cite{Bron} and references there in).
	
Though the scalar multiplet model is self-consistent, it is not the direct way for generalization of a plane monopole to the curved space-time. Contrary to the flat geometry, in a curved space-time scalar multiplets and real vectors are transformed differently. In application to the brane world with two extra dimensions the spontaneous symmetry breaking with a vector order parameter was considered in \cite{PRD}. The comparison of the two approaches showed great advantage of the vector order parameter over the scalar multiplets. The equations were of less order and more simple. Their solutions had additional parametric freedom, that allowed the existence of the brane world without any restrictions of fine-tuning type. Particular derivations in \cite{PRD} were performed for the special case of two extra dimensions, diagonal metric tensor, and dependence on the only one coordinate - the distance from the brane.
	
In view of the evident advantages of the vector order parameter, and having in mind its numerous possible applications, it is worth deriving general covariant equations describing gravitational properties of topological defects. It is the main goal of this paper. The result is summarized in section \ref{sec6}. The abilities of the derived equations are demonstrated in application to the brane world concept. In particular, new solutions of the Einstein equations with a transverse vector order parameter are presented in section \ref{sec7:level1}. In the vicinity of phase transition the solutions are found analytically.

\section{\label{sec2}Symmetric and antisymmetric derivative of the vector order parameter}

Let $\phi _{I}$ be a vector order parameter. Its covariant derivative $\phi _{I;K}$ can be presented as a sum of a symmetric $G_{IK}$ and an antisymmetric $F_{IK}$\ terms:
\begin{equation}
\phi _{I;K}=G_{IK}+F_{IK},\qquad G_{IK}=\frac{1}{2}\left( \phi _{I;K}+\phi _{K;I}\right),\qquad F_{IK}=\frac{1}{2}\left( \phi _{I;K}-\phi _{K;I}\right) . \label{G_IK and F_IK}
\end{equation}
In General Relativity the order parameter enters the Lagrangian via scalar bilinear combinations of its covariant derivatives and via a scalar potential $V$ allowing the spontaneous symmetry breaking. A bilinear combination of the covariant derivatives is a 4-index tensor
\begin{equation*}
S_{IKLM}=\phi _{I;K}\phi _{L;M}.
\end{equation*}
The most general form of the scalar $S$, formed via contractions of $S_{IKLM},$ is
\begin{equation}
S=\left( ag^{IK}g^{LM}+bg^{IL}g^{KM}+cg^{IM}g^{KL}\right) S_{IKLM},
\label{Scalar S}
\end{equation}
where $a,b,$ and $c$ are arbitrary constants. The classification of topological defects with vector order parameters is most convenient in terms of the symmetric and antisymmetric parts of $\phi _{I;K}.$ In view of
\begin{equation*}
G_{K}^{L}F_{L}^{K}=0,
\end{equation*}
and
\begin{equation*}
F_{M;K}F^{K;M}=-F_{M;K}F^{M;K}
\end{equation*}
the scalar (\ref{Scalar S}) can be presented in the form
\begin{equation}
S=a\left( G_{K}^{K}\right) ^{2}+\left( b+c\right) G_{K}^{L}G_{L}^{K}+\left( b-c\right) F_{K}^{L}F_{L}^{K}.
\label{S=}
\end{equation}
The last term with antisymmetric derivatives is identical to electromagnetism. It becomes clear in the common notations $A_{I}=\phi _{I}/2,$ $\ F_{IK}=A_{I;K}-A_{K;I}.$ The bilinear combination of the derivatives $F_{IK}F^{IK}$ is the same as in electrodynamics. In view of the symmetry of Christoffel symbols $\Gamma _{IK}^{L}=\Gamma _{KI}^{L},$
\begin{equation*}
A_{I;K}-A_{K;I}=\frac{\partial A_{I}}{\partial x^{K}}-\frac{\partial A_{K}}{\partial x^{I}},
\end{equation*}
and the combination $F_{IK}F^{IK}$ is free of the derivatives of the metric tensor. On the contrary, the two first terms in (\ref{S=}) with symmetric covariant derivatives contain not only the components of the metric tensor $g^{IK},$ but also the derivatives $\frac{\partial g_{IK}}{\partial x^{L}}.$ The difference between the two terms with symmetric tensors is caused by the curvature of space-time, see (\ref{Difference of A and B}) below.

\section{\label{sec3}Lagrangian}

In the notations
\begin{equation}
a=A,\qquad b+c=B,\qquad b-c=C
\label{A,B, and C}
\end{equation}
the Lagrangian is
\begin{equation}
L\left( \phi _{I},\frac{\partial \phi _{I}}{\partial x^{K}},g^{IK},\frac{\partial g_{IK}}{\partial x^{L}}\right) =A\left( G_{M}^{M}\right) ^{2}+BG_{MN}G_{\text{ \ \ \ }}^{MN}+CF_{MN}F_{\text{ \ \ \ }}^{MN}-V\left( \phi _{M}\phi ^{M}\right) .
\label{Lagrangean}
\end{equation}
Dealing with variational derivatives, it is convenient to express all terms in (\ref{Lagrangean}) as functions of $\phi _{I},\frac{\partial \phi _{I}}{\partial x^{K}},g^{IK},\frac{\partial g_{IK}}{\partial x^{L}}$:
\begin{equation*}
\phi ^{K}=g^{IK}\phi _{I},\ \ \phi _{I;K}=\frac{\partial \phi _{I}}{\partial x^{K}}-\Gamma _{IK}^{L}\phi _{L},\ \ \ \Gamma _{IK}^{L}=\frac{1}{2}g^{LM}\left( \frac{\partial g_{MI}}{\partial x^{K}}+\frac{\partial g_{MK}}{\partial x^{I}}-\frac{\partial g_{IK}}{\partial x^{M}}\right) ,....
\end{equation*}

\section{\label{sec4}Vector field equations}
	
The vector field $\phi _{I}$ obeys the Eiler-Lagrange equations
\begin{equation}
\frac{1}{\sqrt{-g}}\frac{\partial }{\partial x^{L}}\left( \sqrt{-g}\frac{\partial L}{\partial \frac{\partial \phi _{I}}{\partial x^{L}}}\right) =\frac{\partial L}{\partial \phi _{I}}.
\label{Eiler-Lagrange equation}
\end{equation}
Here and below $\sqrt{-g}$ stands for $\sqrt{\left( -1\right) ^{D-1}g},$ where $D$ is the dimension of space-time. In terms of $a,b,$ and $c$ the variational derivative $\frac{\partial L}{\partial \frac{\partial \phi _{I}}{\partial x^{L}}}$ is
\begin{equation*}
\frac{\partial L}{\partial \frac{\partial \phi _{I}}{\partial x^{L}}}=2\left( ag^{IL}\phi _{;K}^{K}+b\phi ^{I;L}+c\phi ^{L;I}\right) .
\end{equation*}
For $\frac{\partial L}{\partial \phi _{I}}$ we have
\begin{equation*}
\frac{\partial L}{\partial \phi _{I}}=\left( ag^{NK}g^{LM}+bg^{NL}g^{KM}+cg^{NM}g^{KL}\right) \frac{\partial }{\partial \phi _{I}}\phi _{N;K}\phi _{L;M}-\frac{\partial }{\partial \phi _{I}}V\left( g^{NK}\phi _{N}\phi _{K}\right) .
\end{equation*}
In a locally geodesic reference system (where the Christoffel's symbols together with the derivatives $\frac{\partial g_{IK}}{\partial x^{L}}$ are zeros) $\frac{\partial }{\partial \phi _{I}}\phi _{N;K}\phi _{L;M}=0,$ and $\frac{\partial L}{\partial \phi _{I}}=-2V^{\prime }\phi ^{I}.$
Here
\begin{equation}
V^{\prime }=\frac{dV}{d\left( \phi _{L}\phi ^{L}\right) }.
\label{Definition V'}
\end{equation}
The vector field equations (\ref{Eiler-Lagrange equation}), having a covariant form
\begin{equation}
a\phi _{;L;I}^{L}+b\phi _{I;L}^{;L}+c\phi _{;I;L}^{L}=-V^{\prime }\phi _{I}
\label{Field eq in terms of a,b,c}
\end{equation}
in a locally geodesic system, remain the same in all other reference frames. In terms of $A,$ $B,$ and $C$
\begin{equation}
AG_{L;I}^{L}+BG_{I;L}^{L}-CF_{\text{ \ }I;L}^{L}=-V^{\prime }\phi _{I}.
\label{Field eq in terms of A,B,C}
\end{equation}

There are two independent terms with the symmetric tensor $G$ in (\ref{Field eq in terms of A,B,C})\ and one with the antisymmetric tensor $F.$ The physical difference between the two symmetric terms becomes clear if we set $F_{IK}=0$. Then (\ref{Field eq in terms of A,B,C}) reduces to
\begin{equation}
A\phi _{;L;I}^{L}+B\phi _{;I;L}^{L}=-V^{\prime }\phi _{I}
\label{Field eq in terms of A,B}
\end{equation}
and the two left terms differ by the order of differentiation. In General Relativity the second covariant derivatives are not invariant against the replacement of the order of differentiation:
\begin{equation}
A_{;K;L}^{I}-A_{;L;K}^{I}=-R_{\text{ }MKL}^{I}A^{M}.
\label{Changing order of diff}
\end{equation}
The difference between the two terms in (\ref{Field eq in terms of A,B}) \ is caused by the curvature of space-time:
\begin{equation}
\phi _{;I;L}^{L}-\phi _{;L;I}^{L}=-R_{\text{ }KIL}^{L}\phi ^{K}=R_{IK}\phi ^{K}.
\label{Difference of A and B}
\end{equation}
In the flat space-time the Ricci tensor $R_{IK}=R_{KLI}^{L}=-R_{KIL}^{L}=0,$ and in case $F_{IK}=0$ there is no physical difference between the two first terms in (\ref{Field eq in terms of a,b,c}):
\begin{equation*}
a\phi _{;L;I}^{L}+b\phi _{I;L}^{;L}=\left( a+b\right) \phi _{I;L}^{;L},\qquad R_{IK}=0.
\end{equation*}

\section{\label{sec5}Energy-momentum tensor}
	
In a locally geodesic coordinate system the general expression for the energy-momentum tensor (see a footnote in \cite{PRD})
\begin{equation}
T_{IK}=\frac{2}{\sqrt{-g}}\left[ \frac{\partial \sqrt{-g}L}{\partial g^{IK}}+g_{MI}g_{NK}\frac{\partial }{\partial x^{L}}\left( \sqrt{-g}\frac{\partial L}{\partial \frac{\partial g_{MN}}{\partial x^{L}}}\right) \right]
\label{Tik general}
\end{equation}
can be written in a covariant form as follows:
\begin{equation}
T_{IK}=-g_{IK}L+2\frac{\partial L}{\partial g^{IK}}+2g_{MI}g_{NK}\left( \frac{\partial L}{\partial \frac{\partial g_{MN}}{\partial x^{L}}}\right) _{;L}.
\label{Tik covariant}
\end{equation}
Here we used the identity
\begin{equation*}
\frac{2}{\sqrt{-g}}\frac{\partial \sqrt{-g}}{\partial g^{IK}}=-g_{IK}.
\end{equation*}

	In terms of symmetric and antisymmetric tensors (\ref{G_IK and F_IK}) we find:
\begin{equation}
\frac{\partial L}{\partial g^{IK}}=2\left( AG_{L}^{L}G_{IK}+BG_{K}^{L}G_{IL}+CF_{I}^{\text{ \ \ }L}F_{KL}\right) -V^{\prime }\phi _{I}\phi _{K},
\label{dL/dg=}
\end{equation}
\begin{equation}
\frac{\partial L}{\partial \frac{\partial g_{MN}}{\partial x^{L}}}=-A\phi _{;P}^{P}\left( g^{LN}\phi ^{M}+g^{LM}\phi ^{N}-g^{NM}\phi ^{L}\right) -B\left( G^{LN}\phi ^{M}+G^{LM}\phi ^{N}-G^{MN}\phi ^{L}\right) .
\label{dL/d(dg/dx)=}
\end{equation}
The tensor (\ref{dL/d(dg/dx)=}) is presented in a symmetric form against the indexes $M,N.$
	
Substituting (\ref{dL/dg=}) and (\ref{dL/d(dg/dx)=}) into (\ref{Tik covariant}), we find the following covariant expression for the energy-momentum tensor:
\begin{equation}
\begin{array}{l} T_{IK}=-g_{IK}L+2V^{\prime }\phi _{K}\phi _{I}+2Ag_{IK}\left( G_{M}^{M}\phi ^{L}\right) _{;L}+2B\left[ \left( G_{IK}\phi ^{L}\right) _{;L}-G_{K}^{L}F_{IL}-G_{I}^{L}F_{KL}\right]\\
  +2C\left( 2F_{\text{ }I}^{L}F_{LK}-F_{\text{ \ }K;L}^{L}\phi _{I}-F_{\text{ }I;L}^{L}\phi _{K}\right) .
\end{array}
\label{General Tik Covariant}
\end{equation}

The vector field equations (\ref{Field eq in terms of A,B,C})\ were used to reduce $T_{IK}$ to a rather simple form (\ref{General Tik Covariant}).

\subsection{\label{sec5:level2}Checking the zero of the covariant divergence $T_{I;K}^{K}=0$}

To confirm the correctness of the energy-momentum tensor (\ref{General Tik Covariant}) it is necessary to demonstrate that the covariant divergence

$T_{I;K}^{K}=-L_{;I}+2\left( V^{\prime }\phi ^{K}\phi _{I}\right) _{;K}+2A\left( G_{M}^{M}\phi ^{L}\right) _{;L;I}+2B\left[ \left( G_{IK}\phi ^{L}\right) _{;L}-G^{LK}F_{IL}-G_{I}^{L}F_{\text{ \ \ }L}^{K}\right] _{;K}+2C\left( 2F_{\text{ }I}^{L}F_{L}^{\text{ \ }K}-F_{\text{ \ \ \ };L}^{LK}\phi _{I}-F_{\text{ }I;L}^{L}\phi ^{K}\right) _{;K}$    \\
is zero.

First of all, using the vector field equations (\ref{Field eq in terms of A,B,C}), I exclude $V_{;I}=2V^{\prime }\phi ^{L}\phi _{L;I}$ and $\left( V^{\prime }\phi ^{K}\phi _{I}\right) _{;K}$, and present $T_{I;K}^{K}$ as
\begin{equation}
T_{I;K}^{K}=Aa_{I}+Bb_{I}+Cc_{I}.
\label{Aai+Bbi+Cci}
\end{equation}
Vectors $a_{I},b_{I},$ and $c_{I}$ are too complicated and do not deserve to be written down here. The coefficients $A,B,$ and $C$ are arbitrary constants. However, it doesn't mean that all three vectors in (\ref{Aai+Bbi+Cci}) are zeros separately.

A simple derivation reduces $a_{I}$ to
\begin{equation}
a_{I}=2\left( \phi _{;K;I}^{K}\phi _{L}-\phi _{;K;L}^{K}\phi _{I}\right) ^{;L}.
\label{ai=}
\end{equation}
Reduction of the vectors $b_{I}$ and $c_{I}$ to the similar form
\begin{equation}
b_{I}=2\left( G_{I;K}^{K}\phi _{L}-G_{L;K}^{K}\phi _{I}\right) ^{;L}
\label{bi=}
\end{equation}
\begin{equation}
c_{I}=2\left( F_{\text{ \ }L;K}^{K}\phi _{I}-F_{\text{ \ }I;K}^{K}\phi _{L}\right) ^{;L}
\label{ci=}
\end{equation}
is a more complicated procedure. The identities
\begin{equation}
F_{IL;K}+F_{LK;I}+F_{KI;L}=0
\label{Fik;l+Fkl;i+Fli;k=0}
\end{equation}
\begin{equation}
\left( F_{IL;K}+F_{IK;L}\right) F^{LK}=0
\label{Flk(F+F)=0}
\end{equation}
\begin{equation}
F_{\text{ \ };K;L}^{KL}=0
\label{Fkl;k;l=0}
\end{equation}
were used in deriving (\ref{ci=}). The first one (\ref{Fik;l+Fkl;i+Fli;k=0}) comes out of the definition (\ref{G_IK and F_IK}) of $F_{IK}.$ The second identity (\ref{Flk(F+F)=0}) is a contraction of a symmetric tensor\ (in brackets) with the antisymmetric $F^{LK}$ over the indexes $K,L$. Scalar (\ref{Fkl;k;l=0}) is zero because $F^{KL}$ is an antisymmetric tensor, whereas for any arbitrary tensor $A^{KL}$ the scalar $A_{\text{ \ \ \ };K;L}^{KL}$ is symmetric against the lower indexes:
\begin{equation*}
A_{\text{ \ };L;K}^{LK}=A_{\text{ \ };K;L}^{LK}+g^{QL}R_{TQLK}A^{TK}+g^{PK}R_{TPLK}A^{LT}=A_{\text{ \ };K;L}^{LK}+R_{TK}\left( A^{KT}-A^{TK}\right) =A_{\text{ \ };K;L}^{LK}.
\end{equation*}

Deriving (\ref{bi=}), we changed the order of covariant differentiation
\begin{equation*}
\phi _{L;I;K}=\phi _{L;K;I}+\phi _{P}R_{\text{ \ }LIK}^{P},
\end{equation*}
and used the properties of the Riemann curvature tensor:
 \begin{equation*}
\begin{array}{l}
R_{LPIK}=-R_{LPKI}=-R_{PLIK}=R_{IKLP},\\
R_{LPIK}+R_{LKPI}+R_{LIKP}=0.
\end{array}
\end{equation*}
Also $G_{\text{ \ \ \ }}^{LK}F_{LK;I}=0$ as the contraction of a symmetric and an antisymmetric tensors over the indexes $L,K$.
	
The covariant divergence of the energy-momentum tensor (\ref{Aai+Bbi+Cci}) with $a_{I},b_{I},$ and $c_{I},$ given by (\ref{ai=}-\ref{ci=}), is evidently zero due to the vector field equations (\ref{Field eq in terms of A,B,C}).

\section{\label{sec6}Covariant equations}

If a topological defect, associated with a vector order parameter, plays the dominant role in the formation of the space-time structure, then the gravitational properties of the system are determined by the joint set of vector field equations (\ref{Field eq in terms of A,B,C}) and Einstein equations
\begin{equation}
R_{IK}-\frac{1}{2}g_{IK}R=\varkappa ^{2}T_{IK}
\label{Einstein equation}
\end{equation}
with $T_{IK}$ (\ref{General Tik Covariant}). $\varkappa ^{2}$ is the (multidimensional) gravitational constant. Vector field equations are not independent. They follow from Einstein equations due to Bianchi identities. In practice it is convenient to use a combination of vector field and Einstein equations.

In general there are three arbitrary constants $A,B,$ and $C$. Particular physical cases can be classified by their relations.

\section{\label{sec7}Particular cases. Applications to brane world concept}

From the point of view of macroscopic theory of phase transitions the brane is a topological defect, inevitably accompanying the spontaneous symmetry breaking. The order parameter was traditionally considered as a hedgehog-type multiplet of scalar fields (see \cite{Bron} and references there in) and as a longitudinal spacelike vector ($\phi _{K}\phi ^{K}<0$) \cite{PRD} directed along and depending on the same specified coordinate -- direction from the brane hypersurface. In the case of two extra dimensions the whole $( D=d_{0}+2) $-dimensional space-time has the structure $M^{d_{0}}\times R^{1}\times \Phi ^{1}$ and the metric

\begin{equation}\label{brane metric}
    ds^{2}=e^{2\gamma \left( l\right) }\eta _{\mu \nu }dx^{\mu }dx^{\nu }-\left( dl^{2}+e^{2\beta \left( l\right) }d\varphi ^{2}\right) .
\end{equation}
Here $\eta _{\mu \nu }$ is the flat $d_{0}$-dimensional Minkovsky brane metric $\left( \mu ,\nu =0,1,...,d_{0}-1,\text{ }d_{0}>1\right) $, and $\varphi $ is the angular cylindrical extradimensional coordinate. $\gamma $ and $\beta $ are functions of the distinguished extradimensional coordinate $x^{d_{0}}=l$ -- the distance from the center, i.e. from the brane. $e^{\beta \left( l\right) }=r\left( l\right) $ is the circular radius. Greek indices $\mu ,\nu $,.. correspond to $d_{0}$-dimensional space-time on the brane, and $I,K$,... -- to all $D=d_{0}+2$ coordinates. The metric tensor $g_{IK}$ is diagonal. The curvature of the metric on brane due to the matter is supposed to be much smaller than the curvature of the bulk caused by the brane formation.
	
\subsection{\label{sec7:level1}Symmetric covariant derivative. Longitudinal vector field}

In the case of longitudinal vector field (when $\phi _{I}=\phi \left( l\right) \delta _{Id_{0}}$ is directed along and depends upon the same coordinate $l$), the covariant derivative $\phi _{I;K}$ is a symmetric tensor,
\begin{equation*}
    F_{IK}=0,\ G_{IK}=\phi _{I;K}=\phi _{K;I}.
\end{equation*}
The set of covariant equations
\begin{eqnarray*}
  R_{IK}-\frac{1}{2}g_{IK}R &=&\varkappa ^{2}\left[ 2V^{\prime }\phi _{K}\phi _{I}+2Ag_{IK}\left( \phi _{;M}^{M}\phi ^{L}\right) _{;L}+2B\left( \phi _{I;K}\phi ^{L}\right) _{;L}-g_{IK}L\right] \\
  A\phi _{;K;I}^{K}+B\phi _{;I;K}^{K} &=&-V^{\prime }\phi _{I}
\end{eqnarray*}
contains two free parameters $A$ and $B.$

Earlier the two possibilities $A=\frac{1}{2},B=0$ and $A=0,B=\frac{1}{2}$ were considered in comparison, but separately \cite{PRD}. As compared with the widely used scalar multiplet approach, in the case of the vector order parameter the equations appear to be more simple, while their solutions are more general. In the case of two extra dimensions \cite{PRD} the solutions have one more degree of parametric freedom, in addition to the arbitrary relation between $A$ and $B$. This parametric freedom allows the existence of a brane world without any restrictions of the fine-tuning type.
	
The general covariant equations derived above allow to investigate gravitational properties of more sophisticated longitudinal topological defects with arbitrary relation between $A$ and $B.$ A particular case $A=-B$ is worth attention. In view of (\ref{Difference of A and B}) the difference between the two cases $A$ and $B$ is connected with the curvature of space-time. It vanishes if $R_{IK}=0$. Such an exotic topological defect has no analog in a flat world.

\subsection{\label{sec7:level1}Antisymmetric covariant derivative. Transverse vector field}

In what follows I apply the general covariant equations to the case
\begin{equation}\label{C not eq 0}
    A=B=0,C\neq 0.
\end{equation}
and present a new brane world solution with transverse vector field as the order parameter.
	
In the case (\ref{C not eq 0}) the set of equations (\ref{Field eq in terms of A,B,C}), (\ref{Einstein equation}), and (\ref{General Tik Covariant}) reduces to
\begin{equation}\label{Einst eq antisym}
    R_{IK}-\frac{1}{2}g_{IK}R =\varkappa ^{2}\left[ C\left( 4F_{\text{ }I}^{L}F_{LK}-g_{IK}F_{MN}F_{\text{ \ \ \ }}^{MN}\right) -2V^{\prime }\phi _{K}\phi _{I}+g_{IK}V\right] ,
\end{equation}
\begin{equation}\label{Maxwell equ with V'}
    CF_{\text{ \ }I;L}^{L} =V^{\prime }\phi _{I}.
\end{equation}
	
Antisymmetric tensor $F_{IK}$ is invariant against adding a gradient $\psi _{;I}$ of any scalar $\psi $ to $\phi _{I}$ (gauge invariance):
\begin{equation*}
    \widetilde{F}_{IK}=F_{IK}+\frac{1}{2}\left( \psi _{;I;K}-\psi _{;K;I}\right) =F_{IK}-\frac{1}{2}\left( \Gamma _{IK}^{L}-\Gamma _{KI}^{L}\right) \psi _{;L}=F_{IK}.
\end{equation*}

The set (\ref{Einst eq antisym})-(\ref{Maxwell equ with V'}) is not gauge invariant because the order parameter (``vector-potential'') $\phi _{I}$ enters the equations directly, and not only via the covariant derivatives (``forces'') $F_{IK}.$ Only if $V=V_{0}=const,$ $V^{\prime }=0$ the equations (\ref{Einst eq antisym})-(\ref{Maxwell equ with V'}) reduce to the gauge invariant set of Einstein-Maxwell equations (with no charges). In this case the term $\varkappa ^{2}g_{IK}V_{0}$ plays the role of a cosmological constant $\Lambda =\varkappa ^{2}V_{0}$.
	
In the space-time with metric (\ref{brane metric})
\begin{equation}\label{F_ik =}
    F_{IK}=\frac{1}{2}\left( \delta _{Kd_{0}}\phi _{I}^{\prime }-\delta _{Id_{0}}\phi _{K}^{\prime }\right) .
\end{equation}

 For a longitudinal vector $\phi _{I}=\phi \delta _{Id_{0}}$ the antisymmetric covariant derivative (\ref{F_ik =}) is zero. A nonzero $F_{IK}$  exists if $\phi _{I}$ is a transverse vector. It should be directed along and depend on different coordinates. In the case of two extra dimensions $\left( x^{d_{0}},\varphi \right) $ the vector $\phi _{I}$ can be directed azimuthally,
\begin{equation}\label{transverse field}
    \phi _{I}=\phi \delta _{I\varphi },
\end{equation}
provided that all functions depend on $x^{d_{0}}=l.$
	
The Einstein equations (\ref{Einst eq antisym}), written in the form $R_{IK}=\varkappa ^{2}\left( T_{IK}-\frac{1}{d_{0}}g_{IK}T\right) $,\ are
\begin{eqnarray} \label{Antisym set}
\nonumber 
  \gamma ^{\prime \prime }+\gamma ^{\prime }\left( d_{0}\gamma ^{\prime }+\beta ^{\prime }\right) =-\frac{\varkappa ^{2}}{d_{0}}\left( Ce^{-2\beta }\phi ^{\prime 2}+2V+2V^{\prime }e^{-2\beta }\phi ^{2}\right)  \\
  -\left( d_{0}\gamma ^{\prime \prime }+\beta ^{\prime \prime }+d_{0}\gamma ^{\prime 2}+\beta ^{\prime 2}\right) =\frac{\varkappa ^{2}}{d_{0}}\left[ -C\left( d_{0}-1\right) e^{-2\beta }\phi ^{\prime 2}+2V+2V^{\prime }e^{-2\beta }\phi ^{2}\right]  \\
  \nonumber
  \beta ^{\prime \prime }+\beta ^{\prime }\left( d_{0}\gamma ^{\prime }+\beta ^{\prime }\right) =\frac{\varkappa ^{2}}{d_{0}}\left[ C\left( d_{0}-1\right) e^{-2\beta }\phi ^{\prime 2}-2V+2V^{\prime }\left( d_{0}-1\right) e^{-2\beta }\phi ^{2}\right] .
\end{eqnarray}

The prime $^{\prime }$ denotes the derivative $\frac{d}{dl}$ $:$ $\ \gamma ^{\prime }=\frac{d\gamma }{dl},\ \beta ^{\prime }=\frac{d\beta }{dl},\ \phi ^{\prime }=\frac{d\phi }{dl},$ \ except $V^{\prime }=\frac{dV}{d\left( \phi _{K}\phi ^{K}\right) }$ (\ref{Definition V'}). $\gamma \left( l\right) $  enters the set (\ref{Antisym set})  only via the derivatives $\gamma ^{\prime }$ and $\gamma ^{\prime \prime }.$ The set (\ref{Antisym set}) is of the fourth order with respect to unknowns $\gamma ^{\prime },\beta ,$ and $\phi .$ The independent variable $l$ is a cyclic coordinate. Excluding the second derivatives $\beta ^{\prime \prime }$ and $\gamma ^{\prime \prime },$ we get a relation
\begin{equation}\label{first int}
    \gamma ^{\prime }\beta ^{\prime }+\varkappa ^{2}\left( \frac{C}{2d_{0}}e^{-2\beta }\phi ^{\prime 2}+\frac{V}{d_{0}}\right) +\frac{d_{0}-1}{2}\gamma ^{\prime 2}=0,
\end{equation}
which can be considered as a first integral of the system (\ref{Antisym set}).
	
The vector field equation (\ref{Maxwell equ with V'}) reduces to
\begin{equation}\label{fi''+()}
    \phi ^{\prime \prime }+\left( d_{0}\gamma ^{\prime }-\beta ^{\prime }\right) \phi ^{\prime }=\frac{2}{C}V^{\prime }\phi .
\end{equation}

If the influence of matter on the brane is neglected, then there is no physical reason for singularities. For the metric (\ref{brane metric}) the Riemann curvature tensor is regular if the combinations
\begin{equation*}
    \gamma ^{\prime },\qquad \gamma ^{\prime \prime }+\gamma ^{\prime 2},\qquad \beta ^{\prime }\gamma ^{\prime },\qquad \beta ^{\prime \prime }+\beta ^{\prime 2}
\end{equation*}
are finite\cite{Bron}. The finiteness of $\beta ^{\prime \prime }+\beta ^{\prime 2}$ ensures the correct $\left( =2\pi \right) $ circumference-to-radius ratio at $l\rightarrow 0$ . It excludes the curvature singularity in the center $l=0,$ which is a singular point of the cylindrical coordinate system. At $l\rightarrow 0$ we have $\beta ^{\prime }=\frac{1}{l}+O\left( l\right) .$ Finiteness of $\beta ^{\prime }\gamma ^{\prime }$ dictates $\gamma ^{\prime }\left( 0\right) =0.$ As far as the left hand sides of the equations (\ref{Antisym set}) are finite, the combinations $e^{-2\beta }\phi ^{\prime 2}=\frac{\phi ^{\prime 2}}{r^{2}}$ and $e^{-2\beta }\phi ^{2}$ $=\frac{\phi ^{2}}{r^{2}}$ should also be finite. Hence, for the transverse vector field at $l\rightarrow 0$ we have
\begin{equation}\label{betta' and others at l to zero}
    \beta ^{\prime }=\frac{1}{l},\qquad \gamma ^{\prime }=\gamma _{0}^{\prime \prime }l,\qquad \phi ^{\prime }=\phi _{0}^{\prime \prime }l,\qquad \phi =\frac{1}{2}\phi _{0}^{\prime \prime }l^{2},\text{ \ \ \ }l\rightarrow 0.
\end{equation}

Only one of the two constants $\gamma _{0}^{\prime \prime }$ and $\phi _{0}^{\prime \prime }$ in (\ref{betta' and others at l to zero}) remains arbitrary. From the first integral (\ref{first int}) we find the connection between $\gamma _{0}^{\prime \prime }$ and $\phi _{0}^{\prime \prime }:$
\begin{equation}\label{relation between a and b}
    \gamma _{0}^{\prime \prime }+\frac{\varkappa ^{2}}{2d_{0}}\left( C\phi _{0}^{\prime \prime 2}+2V_{0}\right) =0.
\end{equation}
Here $V_{0}$ is the value of $V$ at $l=0.$ In view of (\ref{betta' and others at l to zero}) $\phi _{K}\phi ^{K}=\frac{\phi ^{2}}{r^{2}}=0$ at $l=0,$ and $V_{0}=V\left( 0\right) .$ The solutions with the transverse vector field and with the longitudinal one are different. In case of the longitudinal field, see\cite{PRD}, we had $\phi =\phi _{0}^{\prime }l$ at $l\rightarrow 0.$
	
For the regular solutions, spreading over the whole interval $0\leq l<\infty ,$ we find from the Einstein equations (\ref{Antisym set}) the following limiting values $\beta _{\infty }^{\prime }$ and $\gamma _{\infty }^{\prime }$ at $l\rightarrow \infty :$
\begin{equation*}
    \beta _{\infty }^{\prime }=\gamma _{\infty }^{\prime }=\sqrt{-\frac{2\varkappa ^{2}V_{\infty }}{d_{0}\left( d_{0}+1\right) }}.
\end{equation*}
Here $V_{\infty }$ is the limit of $V$ at $l\rightarrow \infty .$ For the solutions with infinitely growing $r\left( l\right) \sim e^{\beta _{\infty }^{\prime }l}$ the scalar $\phi _{K}\phi ^{K}=\frac{\phi ^{2}}{r^{2}}\rightarrow 0$ at $l\rightarrow \infty ,$ and \ $V_{\infty }$ coincides with $V_{0}=V\left( 0\right) .$ The necessary condition of existence of the regular solutions, terminating at $l=\infty ,$ is $V_{\infty }=V_{0}<0.$

If $\phi _{0}^{\prime \prime }=0,$ then the field equation (\ref{fi''+()}) has a trivial solution $\phi \equiv 0,$ corresponding to the state of unbroken symmetry. In the broken symmetry state $\phi _{0}^{\prime \prime }\neq 0$ the order parameter is not zero. However in the close vicinity of the phase transition $\phi \left( l\right) $ is very small, and its influence on the metric is negligible. In the vicinity of phase transition, where $\phi _{0}^{\prime \prime }\ll 1,$ the equations (\ref{Antisym set}) get so simplified
\begin{eqnarray*}
  \nonumber \gamma ^{\prime \prime }+\frac{d_{0}+1}{2}\gamma ^{\prime 2}+\frac{\varkappa ^{2}V_{0}}{d_{0}} =0 \\
  \nonumber \beta ^{\prime }-\left( \ln \gamma ^{\prime }\right) ^{\prime }-\gamma ^{\prime } =0,
\end{eqnarray*}
that for $\gamma ^{\prime }$ and $\beta ^{\prime }$ there is the analytical solution
\begin{equation*}
    \gamma ^{\prime }\left( l\right)  = \gamma _{\infty }^{\prime }\tanh \left( \frac{d_{0}+1}{2}\gamma _{\infty }^{\prime }l\right)
\end{equation*}
\begin{equation}\label{gamma' and betta' near transition}
    \beta ^{\prime }\left( l\right)  = \gamma _{\infty }^{\prime }\left[ \frac{\left( d_{0}+1\right) }{\sinh \left[ \left( d_{0}+1\right) \gamma _{\infty }^{\prime }l\right] }+\tanh \left( \frac{d_{0}+1}{2}\gamma _{\infty }^{\prime }l\right) \right] .
\end{equation}

In the vicinity of phase transition the field $\phi \left( l\right) $ obeys the equation (\ref{fi''+()}) with $\gamma ^{\prime }$ and $\beta ^{\prime }$ from (\ref{gamma' and betta' near transition}) and constant $V^{\prime }=V^{\prime }\left( 0\right) \equiv V_{0}^{\prime }.$ If $V_{0}^{\prime }$ is zero, then  $\phi \left( l\right) $ is a growing function terminating at $\phi \left( \infty \right) =\frac{2\phi _{0}^{\prime \prime }}{\left( d_{0}^{2}-1\right) \gamma _{\infty }^{\prime 2}}:$
\begin{equation*}
    \phi \left( l\right) =\frac{2\phi _{0}^{\prime \prime }}{\left( d_{0}^{2}-1\right) \gamma _{\infty }^{\prime 2}}\left\{ 1-\left[ \cosh \left( \frac{d_{0}+1}{2}\gamma _{\infty }^{\prime }l\right) \right] ^{-2\frac{d_{0}-1}{d_{0}+1}}\right\} ,\quad V_{0}^{\prime }=0.
\end{equation*}

If $V_{0}^{\prime }\neq 0,$ then the asymptotic behavior of $\phi \left( l\right) $\ is determined by the equation
\begin{equation*}
    \phi ^{\prime \prime }+\left( d_{0}-1\right) \gamma _{\infty }^{\prime }\phi ^{\prime }-\frac{2V_{0}^{\prime }}{C}\phi =0.
\end{equation*}
Its regular solutions, oscillating and vanishing at $l\rightarrow \infty ,$ exist only if $V_{0}^{\prime }/C <0.$
	
To avoid overloading the paper with specific details of particular solutions, I intend to publish the complete analytical and numerical analysis of the new brane world solutions with the transverse vector order parameter elsewhere. These solutions are presented here for demonstration of abilities of the general covariant equations -- the main result of this paper.

\end{document}